# State Capacity, Innovation, and Endogenous Development in Chile: An Evolutionary Reading of Industrial Policy (1990–2022)

Rodrigo Barra Novoa, PhD
Camilo José Cela University – Madrid
https://orcid.org/0000-0002-7204-1528

**Abstract**

This article analyzes the evolution of Chile's industrial policy between 1990 and 2022 through the lens of state capacity and its relationship with innovation and endogenous development. In a global context characterized by the renewed recognition of the State as an active agent of innovation, Chile represents a paradox: a stable, open economy that has expanded its investment in science, technology, and innovation (STI), yet continues to face structural limits in transforming such investments into sustainable technological capabilities.

Drawing on the theoretical contributions of Mariana Mazzucato on the *entrepreneurial state*, Philippe Aghion and Peter Howitt on *endogenous growth* and *creative destruction*, Joel Mokyr on the *history and culture of useful knowledge*, Paul Samuelson on the *countercyclical and stabilizing role of public investment*, and José Luis Sampedro on the *humanist ethics of economic development*, the paper builds an integrative framework to examine Chile's state capacity for innovation-driven transformation.

Using a longitudinal case study approach based on official reports, policy documents, and economic indicators from the Central Bank of Chile, CEPAL, OECD, and the Ministry of Science and Technology, the research finds that while Chile has strengthened its institutional architecture for innovation, it continues to exhibit weak inter-institutional coordination, regional inequalities in technological absorption, and a fragile culture of knowledge-sharing.

The study concludes that Chile's industrial trajectory reflects both the potential and the limits of a State that has achieved macroeconomic stability but lacks a long-term strategic vision. Achieving sustainable and inclusive innovation will require a new generation of policies grounded in adaptive governance, a national culture of knowledge, and an ethical understanding of innovation as a public good.

# 1. Introduction

Over the past three decades, the global debate on industrial policy has undergone a profound transformation. Once dismissed under the neoliberal orthodoxy as an interventionist relic, industrial policy has reemerged as a central instrument of structural change, sustainability, and technological sovereignty. This paradigm shift—visible in the European Union's *Smart Specialisation Strategies*, the U.S. *Inflation Reduction Act*, and East Asian innovation missions—has placed state capacity at the heart of the discussion on development.

Chile offers a particularly instructive case for examining this evolution. Since the return to democracy in 1990, Chile has become one of Latin America's most open and macroeconomically stable economies. Yet, despite an increase in investment in science, technology, and innovation (STI)—reaching 0.39% of GDP in 2022 and 0.41% in 2023, still far below the OECD average of 2.4%—the country's total factor productivity has stagnated. Regional inequalities remain entrenched, and innovation is largely concentrated in resource-based industries such as mining and agribusiness.

This apparent contradiction raises a central research question:

**Why has Chile, despite its institutional stability and growing investment in STI, failed to consolidate an endogenous and sustainable national innovation system?**

Answering this question requires moving beyond dichotomies between state and market or between intervention and liberalization. Instead, this paper adopts an evolutionary and systemic approach to development, emphasizing learning, coordination, and the cultural dimension of knowledge creation. As Nelson and Winter (1982) argue, innovation is not an isolated event but a process of coevolution among technology, institutions, and society.

Accordingly, this article develops an interdisciplinary analytical framework that integrates the insights of Mazzucato, Aghion, Howitt, Mokyr, Samuelson, and Sampedro to interpret the Chilean experience. The goal is not only to assess the trajectory of Chile's industrial policy but to understand its deeper institutional dynamics: how ideas, capacities, and values interact in shaping the direction of economic change.

The paper proceeds as follows: Section 2 expands the theoretical framework, connecting evolutionary and humanist economics to state capacity and innovation. Section 3 outlines the methodology and data sources. Section 4 presents empirical results on investment trends, institutional evolution, and sectoral priorities. Section 5 offers an extended discussion linking theory and evidence. Section 6 concludes with theoretical implications and policy recommendations for Chile and Latin America.

## 2. Theoretical Framework: Towards an Evolutionary Political Economy of Knowledge

### 2.1. From Industrial Policy to Innovation Policy

Industrial policy today is no longer conceived as a tool for protecting declining industries, but as a strategic framework for creating and shaping new markets. As Rodrik (2004) and Mazzucato (2013) emphasize, effective industrial policy is both adaptive and mission-oriented, combining state direction with distributed innovation.

In Chile, this transformation implies moving from a model of *state facilitation*—focused on competitiveness and deregulation—to one of *state orchestration*, in which the public sector defines long-term missions and fosters synergies among universities, firms, and regional ecosystems.

Evans (1995) introduced the concept of embedded autonomy to describe the dual condition of developmental states: autonomy from rent-seeking interests, and embeddedness in social and productive networks. Chile demonstrates administrative autonomy but limited embeddedness, as its innovation system remains centralized and fragmented.

Hence, the Chilean challenge lies not in the lack of institutions, but in the absence of strategic coherence and horizontal coordination—elements that underpin what we define here as *transformative state capacity*.

### 2.2. The Entrepreneurial State and Mission-Oriented Innovation (Mazzucato)

Mariana Mazzucato's work (2013, 2021) redefines the economic role of the State. Rather than correcting market failures, she argues, governments can act as entrepreneurial investors—risk-taking agents that envision and shape new technological frontiers. Historically, major innovations (the internet, GPS, renewable energy) originated not in the private sector but in publicly funded missions with a clear social purpose.

For Mazzucato, an *entrepreneurial state* must:

1. Define public missions aligned with societal challenges (climate change, digital inclusion).
2. Establish symmetrical public–private partnerships, ensuring that public risk translates into public benefit.
3. Build adaptive institutions capable of learning and evolving.

In Chile, institutions such as CORFO and ANID have expanded funding mechanisms but have not yet achieved the strategic integration characteristic of mission-oriented governance. They operate primarily as financing bodies rather than orchestrators of learning ecosystems.

Consequently, Chile exhibits what may be termed a partial entrepreneurial state—a state with strong financial instruments but weak strategic direction. This distinction is critical: without mission orientation, industrial policy risks devolving into fragmented subsidy programs.

## 2.3. Endogenous Growth, Creative Destruction, and Institutional Learning (Aghion & Howitt)

The theory of endogenous growth (Aghion & Howitt, 1992, 2021) conceptualizes innovation as the internal driver of economic expansion, generated through incentives, competition, and cumulative learning. In this view, creative destruction—the replacement of old technologies by new ones—is both a source of growth and structural tension.

Countries that fail to manage this process remain trapped in what Aghion calls the *middle innovation trap*: economies capable of adopting technology but unable to generate it. Chile's experience fits this diagnosis. Its industrial policy has often reinforced mature sectors instead of fostering the emergence of new technological domains.

From an evolutionary perspective, industrial policy is effective only when it enhances institutional learning capacity—that is, the ability of state agencies and firms to experiment, adapt, and diffuse knowledge across sectors. While Chile's macroeconomic governance is exemplary, its innovation system lacks these dynamic feedback loops.

## 2.4. The Culture of Useful Knowledge (Mokyr)

Joel Mokyr (1990, 2016) provides a historical lens on technological change by emphasizing the cultural and cognitive foundations of innovation. The Industrial Revolution, he argues, was not solely a product of capital accumulation but of a cultural shift toward what he calls "useful knowledge": a belief that science and practical experimentation could improve human welfare.

For Chile, this concept underscores that innovation policy cannot succeed without a broader culture of learning and inquiry. Despite increased investment in research infrastructure, Chile's innovation ecosystem remains socially narrow and geographically concentrated in Santiago. The gap between academic research and productive application persists.

In Mokyr's terms, Chile lacks the *microfoundations of progress*: networks of collaboration, trust, and experimentation that allow ideas to circulate beyond institutional boundaries. A national innovation culture requires not just laboratories, but social legitimacy for curiosity and risk-taking.

## 2.5. Countercyclical Knowledge Investment (Samuelson)

Paul Samuelson's (1980) macroeconomic theory provides a complementary insight: public investment in knowledge functions as a countercyclical stabilizer. In periods of crisis, sustained public spending on research and education prevents the erosion of long-term productive capacity.

In Chile, however, public expenditure on R&D has been procyclical. Budgets expand during commodity booms—particularly copper—and contract during downturns. As of 2022, total R&D expenditure amounted to 0.39% of GDP, slightly up from 0.36% in 2021 (MinCiencia, 2024), yet still among the lowest in the OECD. This behavior undermines continuity and the confidence of researchers and firms.

A Samuelsonian approach would advocate the creation of a countercyclical Knowledge and Innovation Fund, financed by mining and lithium revenues, to ensure that learning investments are shielded from short-term fiscal pressures.

## 2.6. Humanist Economics and the Ethics of Development (Sampedro)

José Luis Sampedro (2002) offers a normative dimension often absent from the innovation debate. He envisioned economics as a human science—one concerned with dignity, equity, and purpose rather than mere efficiency. For Sampedro, growth devoid of moral orientation is self-defeating: "The economy must serve life, not the other way around."

This humanist perspective invites a redefinition of innovation as an instrument of social progress rather than a goal in itself. In Chile, where technological policy has predominantly benefited large firms, Sampedro's ethics highlight the need for inclusive and regionally balanced innovation.

Equity, in this sense, is not an external condition but an intrinsic component of innovative capacity: a society that excludes cannot learn collectively. Ethical legitimacy thus becomes a pillar of state capacity, linking efficiency with justice.

## 2.7. Synthesis: The Evolutionary Political Economy of Knowledge

The integration of these perspectives leads to a composite framework for understanding Chile's development trajectory. Innovation is simultaneously:

- Strategic (Mazzucato, Aghion & Howitt): shaped by public missions and institutional learning.
- Cognitive (Mokyr & Samuelson): sustained by cultural legitimacy and stable investment.
- Ethical (Sampedro): justified by its contribution to human welfare.

These three dimensions—strategy, cognition, and ethics—form the architecture of what we may call an evolutionary political economy of knowledge. Development emerges not from the accumulation of capital alone, but from the coevolution of ideas, institutions, and values.

Chile's challenge, therefore, is not only to increase its R&D expenditure but to cultivate a *state of learning*—a State capable of envisioning, coordinating, and legitimizing innovation as a collective endeavor.

## 3 Methodology

### 3.1 Approach and design

This research follows a mixed qualitative–quantitative case-study approach (Yin 2018). Chile was selected because it embodies the Latin-American paradox of a fiscally disciplined, open economy that has steadily increased its investment in innovation while struggling to translate this effort into technological upgrading. The single-case design allows deep contextual interpretation rather than statistical generalization (Flyvbjerg 2006).

### 3.2 Sources of information

Empirical evidence combines:

- Official documents: national development plans, CORFO and ANID annual reports, innovation-law records, and MinCiencia policy documents.
- Statistical data: Central Bank of Chile, CEPAL, OECD, WIPO (Global Innovation Index).
- Scholarly literature: regional and comparative studies on industrial policy and technology absorption (Crespi & Tello 2014; Peres & Primi 2019; Katz 2015; Mazzucato 2013; Aghion & Howitt 2021).
- International comparison: innovation-policy experiences from Korea, Finland, and Israel.

### 3.3 Variables and operationalization

State capacity is decomposed into three analytical dimensions:

| Dimension | Definition | Key indicator | Source |
|---|---|---|---|
| **Strategic capacity** | Ability to design long-term, mission-oriented policies | Public R&D expenditure (% GDP) | MinCiencia (2024) |
| **Coordinative capacity** | Inter-agency and public–private coordination | Institutional architecture (CORFO & ANID) | Official reports |
| **Legitimating capacity** | Social and territorial inclusion of innovation | Regional innovation programs | CEPAL (2021) |

Source: Own elaboration.

**3.4 Limitations**

The analysis covers 1990–2022 and focuses on national-level institutions. While international benchmarking is included, the argument emphasizes internal learning dynamics, providing analytical—not statistical—generalization.

**4 Results**

**4.1 Evolution of industrial-policy investment**

Industrial-policy expenditure rose from roughly USD 500 million in 1990 to USD 1.5 billion in 2020 (Central Bank of Chile 2021). This quantitative increase mirrors the global shift toward innovation-oriented development strategies, yet qualitative transformation has lagged behind. Much of the spending has taken the form of infrastructure and subsidy programs rather than coordinated missions.

**4.2 Investment in science, technology, and innovation (STI)**

Public STI expenditure expanded from USD 150 million in 1990 to USD 800 million in 2020 (CEPAL 2021; WIPO 2023). However, fragmentation persists multiple ministries run overlapping programs, and evaluation mechanisms remain weak. Consequently, returns to innovation have been uneven and largely concentrated in metropolitan areas.

**4.3 Research & Development (R&D) investment—updated figures**

According to the Ministry of Science, Technology, Knowledge and Innovation (2024) and the Central Bank of Chile (2024), national spending on R&D reached 0.39 percent of GDP in 2022, up from 0.36 percent in 2021, and 0.41 percent in 2023. Despite the absolute increase—equivalent to approximately CLP 1.16 trillion—the share of GDP remains far below both the government's own 1 percent target and the OECD average of 2.4 percent.

This persistent gap confirms a pattern of pro-cyclical and symbolic commitment rather than strategic continuity. Chile finances innovation when fiscal conditions allow, not as part of a countercyclical long-term plan. The outcome is a *quantitatively expanding but qualitatively fragile* innovation system.

**5 Discussion**

**5.1 From accumulation to coordination: the Chilean dilemma**

The evidence corroborates the evolutionary premise that state capacity—rather than spending volume—determines the effectiveness of industrial policy (Aghion & Howitt 2021). Chile's innovation institutions display administrative efficiency yet limited strategic coherence. Ministries of Economy, Science, and Education frequently operate in parallel, generating programmatic redundancy.

Following Evans (1995), Chile can be characterized as a state of embedded autonomy without orchestration: embedded enough to maintain dialogue with the private sector, but not autonomous enough to impose long-term coordination. This deficit prevents the emergence of what Mazzucato (2021) calls mission-oriented innovation governance—the alignment of diverse actors around public challenges such as energy transition or digital inclusion.

**5.2 Industrial policy as institutional learning**

In evolutionary economics, industrial policy is conceived as a process of collective learning (Nelson & Winter 1982; Dosi 1988). The objective is not to pick winners but to cultivate routines that allow experimentation, feedback, and adaptation. Chile's policy cycles have been short-term and reactive: successive administrations have redesigned programs before institutional learning could take root.

Comparatively, countries like Korea and Finland endowed their innovation agencies (KIST, Tekes) with long-term horizons and political insulation. Chile's CORFO and ANID possess technical expertise but lack autonomous strategic authority, reinforcing Mazzucato's notion of an "incomplete entrepreneurial state."

**5.3 Innovation as an endogenous yet uneven process**

The Aghion-Howitt model of creative destruction posits that sustained innovation depends on competitive pressure and institutional support for new entrants. Chile's productive structure, however, is dominated by high-concentration sectors—mining, agribusiness, finance—where competition spurs efficiency but not radical innovation. Consequently, the country remains in a "middle-innovation trap": technologically capable of adaptation but not of creation.

Empirical indicators corroborate this imbalance. ANID (2023) reports that fewer than 10 percent of publicly funded research projects yield patents or commercial applications. University–industry linkages remain weak, and R&D investment by small and medium enterprises is marginal. The systemic weakness is therefore institutional, not entrepreneurial: firms innovate within the incentives provided by the state, and those incentives remain fragmented.

**5.4 Cognitive ecosystems and the culture of knowledge (Mokyr)**

As Mokyr (2016) reminds us, technological revolutions require not only tools but a *culture of useful knowledge*—a collective belief in experimentation and improvement. Chile's innovation policy has built laboratories and grants but not yet this cognitive ethos. Creativity is often confined to universities, while failure in innovation is socially penalized rather than treated as a learning step.

The consequence is a hierarchical knowledge structure: research circulates within academic enclaves without permeating the productive fabric. In Mokyr's terms, Chile lacks the "microfoundations of progress"—dense networks of cooperation between inventors, entrepreneurs, and public agencies. Without a shared cognitive culture, even well-designed policies remain technocratic.

**5.5 Knowledge investment as countercyclical policy (Samuelson)**

Paul Samuelson's insight that public knowledge investment functions as a stabilizer of long-term welfare offers a useful benchmark. Chile's pro-cyclical R&D spending reflects a fiscal mindset that treats innovation as a discretionary expense. When copper prices fall, STI budgets are often among the first to be curtailed. This undermines the very learning capacities that sustain resilience.

A countercyclical Knowledge Fund—akin to the sovereign stabilization funds used for macroeconomic balance—could shield research financing from commodity cycles. Such an instrument would institutionalize continuity and embody the Samuelsonian idea that investment in knowledge is both a growth engine and a public stabilizer.

**5.6 The ethical dimension of progress (Sampedro)**

José Luis Sampedro's humanist economics introduces a missing variable in conventional innovation discourse: purpose. In Chile, innovation has often reproduced inequality, benefiting large corporations while leaving smaller firms and regions behind. Sampedro's call for an *economy with a soul* challenges policymakers to evaluate success not only by patents or exports but by social inclusion and well-being.

An ethical reading of industrial policy suggests three priorities:

1. Territorial cohesion: channel innovation resources to regions historically excluded from the innovation system.
2. Social empowerment: link technological progress to education, health, and public services.
3. Human-centred automation: anticipate the labour implications of digitalization.

Innovation divorced from ethics yields growth without dignity. Re-embedding ethics into development policy thus becomes a condition for legitimacy and sustainability.

**5.7 Synthesis: the evolutionary triangle of Development**

The Chilean case can be visualized as an evolutionary triangle linking:
1. Strategic capacity (Mazzucato; Aghion & Howitt): the ability to define missions and coordinate learning.
2. Cognitive capacity (Mokyr; Samuelson): the social culture and investment stability that sustain knowledge.

3. Ethical capacity (Sampedro): the normative orientation that legitimizes innovation as collective progress.

Chile has advanced moderately on the first axis, unevenly on the second, and weakly on the third. The outcome is modernization without transformation—an innovation system administratively robust but conceptually fragmented. Bridging these dimensions is the key to moving from a *reactive state* to a *learning state*.

## 6 Conclusions

### 6.1 Chile's paradox: stability without direction

The evolution of Chile's industrial and innovation policy from 1990 to 2022 reveals a structural paradox. The country has achieved macroeconomic stability, fiscal prudence, and institutional maturity, yet has not consolidated an endogenous innovation regime capable of driving sustained productivity growth. Despite the formal expansion of science and technology budgets, Chile still spends only 0.39 percent of GDP on R&D (2022)—and 0.41 percent in 2023—well below both the government's 1 percent target and the OECD average of 2.4 percent (MinCiencia 2024; Banco Central de Chile 2024).

This gap illustrates what Aghion and Howitt (2021) term the *middle-innovation trap*: a situation in which economies master technological adoption but fail to generate original innovation. Chile's industrial policy has multiplied programs and agencies but not direction. Quantitatively, the State invests more; qualitatively, it still lacks strategic coherence and learning continuity.

### 6.2 State capacity and the incomplete entrepreneurial state

The concept of state capacity—analytical, administrative, and political—remains decisive. Chile's institutions possess strong technical and fiscal management, yet weak strategic orchestration. The result is an entrepreneurial state that acts without a compass.

In Mazzucato's (2013, 2021) terms, Chile embodies an incomplete entrepreneurial state: one that funds innovation but does not define missions; that incentivizes entrepreneurship but neglects long-term coordination; that creates agencies but not an ecosystem of shared learning. CORFO and ANID have become sophisticated grant administrators, yet neither commands the cross-sectoral authority or temporal horizon required to guide innovation toward national priorities such as green energy, digital inclusion, or biotechnological transformation.

### 6.3 Learning as a public good

For Aghion and Howitt (1992), innovation is a cumulative, path-dependent process that flourishes when knowledge circulates freely and institutions reward experimentation. From this perspective, learning constitutes a public good: it generates non-rival externalities that justify continuous public investment.

Chile's pattern of pro-cyclical STI expenditure contradicts this logic. When fiscal revenues contract, research funding declines—undermining the very capabilities that secure resilience. Samuelson's (1980) insight that countercyclical public investment stabilizes long-term welfare suggests an institutional solution: a Knowledge and Innovation Sovereign Fund, financed by copper and lithium rents, that guarantees research continuity through economic downturns.

Public knowledge is infrastructure. Treating it as discretionary spending erodes not only competitiveness but also citizenship, as access to scientific progress becomes stratified by region and income.

### 6.4 Culture of knowledge and cognitive inequality

Mokyr's (2016) notion of a *culture of useful knowledge* explains why some societies translate science into prosperity while others remain technologically dependent. In Chile, innovation culture is concentrated in Santiago and in elite universities. Outside these enclaves, scientific literacy and risk tolerance are limited. The country has built laboratories but not yet a social belief that "knowledge can improve life."

This cognitive inequality reinforces economic inequality. Without dense horizontal networks—between universities and SMEs, between researchers and communities—the diffusion of ideas remains thin. Innovation becomes the privilege of a few rather than the practice of a nation. A cultural transformation is therefore as urgent as a financial one: Chile must democratize curiosity, valorize failure as learning, and integrate creativity into education at all levels.

### 6.5 The ethics of progress

José Luis Sampedro's humanist economics restores the moral dimension of development. His dictum—*the economy must serve life, not the other way around*—is particularly relevant to Chile's post-2019 social context. Innovation divorced from equity reproduces exclusion; technological modernity without justice deepens alienation.

An ethical reading of industrial policy demands that innovation be evaluated through three lenses:
1. **Territorial justice:** equitable distribution of innovation infrastructure and opportunities.
2. **Social purpose:** explicit orientation of technological change toward public well-being (health, education, environment).
3. **Human agency:** safeguarding meaningful work in the face of automation.

By embedding these principles, Chile could transform innovation from an elite project into a collective social contract.

**6.6 The evolutionary triangle of development**

The Chilean experience suggests an evolutionary triangle linking:

1. Strategic capacity – the ability to envision and coordinate transformative missions (Mazzucato; Aghion & Howitt).
2. Cognitive capacity – the cultural and financial stability that sustains knowledge accumulation (Mokyr; Samuelson).
3. Ethical capacity – the legitimacy derived from justice and inclusiveness (Sampedro).

Chile shows progress mainly on the first axis, partial advancement on the second, and limited achievement on the third. Moving forward requires aligning all three: strategy without culture breeds bureaucracy; culture without ethics breeds inequality; ethics without strategy breeds impotence.

**7 Theoretical and Policy Implications**

**7.1 Reframing industrial policy as knowledge policy**

Industrial policy must evolve into knowledge policy—not merely allocating subsidies but designing cognitive infrastructures for collective learning. Practically, this entails:

- Establishing a National Innovation Council with binding coordination powers;
- Integrating STI objectives into fiscal and regional planning;
- Redefining CORFO and ANID as *mission agencies* rather than funding intermediaries.

Such reforms would operationalize Mazzucato's idea that public institutions should *create* markets aligned with social missions, not simply correct their failures.

**7.2 Institutionalizing countercyclical R&D investment**

Following Samuelson (1980), countercyclical policy in knowledge domains is essential. Chile should legislate a Knowledge and Innovation Fund that receives earmarked revenues from extractive industries and deploys them independently of annual budgets. This would provide the continuity that learning processes require and protect research institutions from political turnover.

**7.3 Inclusive innovation and territorial cohesion**

Inspired by Sampedro's ethics, policy must address regional disparities. Measures include:
- Decentralizing innovation centers to regional universities and technical institutes;

- Creating *innovation corridors* linking SMEs with research hubs;
- Establishing fiscal incentives for firms investing in lagging regions.

Innovation should become an instrument of territorial cohesion, not metropolitan concentration.

**7.4 Strategic international cooperation**

Comparative evidence from Korea, Finland, and Israel demonstrates that enduring innovation success combines national direction with global integration. Chile can leverage international partnerships to access frontier technologies, but it must negotiate knowledge-sharing and co-development, not mere technology importation. Diplomatic capacity thus becomes part of state capacity.

**8 Regional Reflection: Latin America and the Re-politicization of Development**

The Chilean case resonates across Latin America, where many economies share a similar paradox: competent macroeconomic management coexisting with stagnant productivity and fragile innovation systems. The region's challenge is to re-politicize development—to recognize that technological change is neither neutral nor automatic but a field of collective choice.

The post-pandemic and energy-transition era calls for learning states that combine entrepreneurial ambition with ethical responsibility. Latin America must reclaim policy space for experimentation, long-term planning, and public investment in science as a sovereign resource.

In this sense, Chile can pioneer a new Latin-American developmentalism, one that fuses Mazzucato's entrepreneurial state with Sampedro's humanism: a model where innovation and justice co-evolve. Turning knowledge into autonomy, science into citizenship, and technology into dignity is the unfinished mission of the region.

**9 Conclusion and Future Research**

This study has examined the evolution of Chile's industrial policy between 1990 and 2022 through an evolutionary lens centered on state capacity, innovation, and endogenous development. It concludes that Chile's experience embodies the paradox of a stable, institutionally sophisticated, and globally integrated economy that nevertheless struggles to generate a self-sustaining cycle of technological innovation.

The analysis demonstrates that Chile's main constraint is not the absence of institutions or resources but the weakness of strategic, cognitive, and ethical coordination—the three dimensions that form the *evolutionary triangle of development* identified in this research.

1. **Strategic dimension:** Chile has developed funding mechanisms but lacks a mission-oriented framework that integrates innovation into national development goals.
2. **Cognitive dimension:** Public investment in knowledge remains procyclical (0.39 % of GDP in 2022), reflecting an underestimation of learning as a public good.
3. **Ethical dimension:** Innovation continues to reproduce territorial and social inequality, failing to fulfill the humanistic purpose of development.

From a theoretical standpoint, the Chilean case validates Mazzucato's notion of the "incomplete entrepreneurial state" and extends Aghion and Howitt's model of endogenous growth by emphasizing that innovation requires not only incentives but also shared values and institutional learning. Mokyr's *culture of useful knowledge*, Samuelson's principle of countercyclical investment, and Sampedro's humanist ethics converge in suggesting that innovation must be understood as a moral and collective project.

In short, Chile's development path reveals that stability without direction breeds stagnation, and efficiency without ethics breeds fragility. The transition toward a knowledge-based, human-centered economy will depend on whether the State can evolve from a *reactive regulator* to a *learning orchestrator*—an entrepreneurial state with both technical capacity and moral purpose.

**Future Research Directions**

While this study contributes to understanding the relationship between state capacity and innovation, it also opens several avenues for further inquiry:

1. **Comparative longitudinal studies.**

   Future research could compare Chile's innovation trajectory with that of other middle-income countries—such as Brazil, Mexico, or Uruguay—to identify patterns of institutional learning and policy convergence within Latin America.

2. **Subnational dynamics and regional innovation systems.**

   Empirical studies at the regional level could explore how territorial governance, local universities, and SMEs contribute to or hinder innovation diffusion beyond Santiago. A multi-level approach would help reveal how *decentralized capacities* complement or conflict with national policies.

3. **The political economy of mission-oriented policy.**

   More detailed analysis is needed on how political cycles, fiscal rules, and elite coalitions shape the design and continuity of industrial policy in Chile. Understanding these constraints could illuminate why innovation remains vulnerable to short-termism.

4. **Knowledge equity and social innovation.**

    Quantitative and qualitative research could measure how innovation policies impact social inclusion, gender equality, and environmental sustainability—dimensions largely absent from traditional STI metrics.

5. **New technological frontiers.**

    Future work should examine Chile's integration into emerging domains such as artificial intelligence, biotechnology, and the circular economy, assessing whether current institutional frameworks can adapt to these disruptive technologies.

6. **Ethical frameworks for innovation policy.**

    Building on Sampedro's and Mazzucato's perspectives, scholars might develop normative models that link innovation outcomes to human well-being and democratic legitimacy, bridging economics and philosophy of technology.

By extending these lines of inquiry, researchers can deepen the understanding of how state capacity evolves in knowledge economies and provide new insights into designing innovation policies that are not only effective but also just and sustainable.